\documentclass[12pt]{article}
\usepackage[xdvi]{graphicx}
 
\begin{document}  
\newcommand{\todo}[1]{{\em \small {#1}}\marginpar{$\Longleftarrow$}}  
\newcommand{\labell}[1]{\label{#1}\qquad_{#1}} 
 
\rightline{DTP/00/109}  
\rightline{hep-th/0012135}  
\vskip 1cm 
\centerline{\Large \bf Looking for event horizons using UV/IR relations}  
\vskip 1cm  
  
\renewcommand{\thefootnote}{\fnsymbol{footnote}}  
\centerline{\bf  
 James P. Gregory\footnote{J.P.Gregory@durham.ac.uk} and Simon
F. Ross\footnote{S.F.Ross@durham.ac.uk}}   
\vskip .5cm  
\centerline{ \it Centre for Particle Theory, Department of 
Mathematical Sciences}  
\centerline{\it University of Durham, South Road, Durham DH1 3LE, U.K.}  
  
\setcounter{footnote}{0}  
\renewcommand{\thefootnote}{\arabic{footnote}}  
 
  
\begin{abstract}  
  A primary goal in holographic theories of gravity is to study the 
  causal structure of spacetime from the field theory point of view. 
  This is a particularly difficult problem when the spacetime has a 
  non-trivial causal structure, such as a black hole. We attempt to 
  study causality through the UV/IR relation between field theory and 
  spacetime quantities, which encodes information about bulk position. 
  We study the UV/IR relations for charged black hole spacetimes in 
  the AdS/CFT correspondence. We find that the UV/IR relations have a 
  number of interesting features, but find little information about 
  the presence of a horizon in the bulk. The scale of Wilson loops is 
  simply related to radial position, whether there is a horizon or 
  not.  For time-dependent probes, the part of the history near the 
  horizon only effects the late-time behaviour of field theory 
  observables. Static supergravity probes have a finite scale size 
  related to radial position in generic black holes, but there is an 
  interesting logarithmic divergence as the temperature approaches 
  zero. 
\end{abstract}   
  
\section{Introduction}   
  
Resolving the long-standing problems associated with black holes in
quantum gravity seems to require a radical shift in our understanding
of spacetime causal structure. Holographic theories of gravity, such
as the AdS/CFT correspondence~\cite{Maldacena:adscft, Gubser:adscft,
Witten:adscft}, seem to offer such a change in viewpoint. The true,
fundamental causal structure of the theory is the fixed background
causal structure in a $d$-dimensional field theory. The dynamical
spacetime description is supposed to emerge from this underlying field
theory in some approximation. Since spacetime has more dimensions than
the space the field theory lives in, the encoding of information about
the dynamical spacetime should be quite subtle. Understanding how the
spacetime, especially its causal structure, are encoded in the field
theory is one of the main open questions about these models. The aim
of this paper is to see to what extent non-trivial causal structures,
such as a black hole horizon, effect the values of simple gauge theory
observables. We find that the causal structure near an event horizon
does not appear in an obvious way in these observables.
  
In the AdS/CFT correspondence, the introduction of a source probe in 
the bulk will be reflected in a change in one-point functions in the 
field theory~\cite{Banks:dynamics,Bal:dynamics,Bal:probes} 
(higher-point functions are also needed to resolve some probes; see 
e.g.~\cite{Bal:particle,Giddings:shell}). The holographic nature of 
the correspondence is reflected in a UV/IR relation between the radial 
position of the probe and the characteristic scale of the one-point 
function in the field theory~\cite{Susskind:holography,Peet:uvir}. In 
the AdS$_5$/CFT$_4$ case, this can be expressed as a distance/distance 
relationship 
\begin{equation} \label{eq:uvir}  
\delta x_\parallel=\frac{\sqrt{g_{YM}^2 N}}{U} 
\end{equation}  
(i.e., a source at radius $U$ in AdS$_5$ corresponds to perturbing the 
field theory in a region of size $\delta x_\parallel$) 
\cite{Peet:uvir}. Here, $U$ is the radial coordinate in a Poincar\'e 
coordinate system, such that $U \to \infty$ at the boundary of 
AdS. Hence (\ref{eq:uvir}) relates large distances in spacetime (the 
IR) to short distances in the field theory (the UV). 
 
In pure AdS$_5$, this relationship follows from the isometry
$x^i\to\lambda x^i,U \to\lambda^{-1} U$ in the bulk. The UV/IR
relationship has also been studied for more general metrics with
Poincar\'e invariance in the directions parallel to the boundary. It
is used to relate non-trivial solutions of this form to
renormalization group flows in the dual field theory (a huge industry
now; early works
are~\cite{Distler:rg,Girardello:rg,Freedman:rg,Klebanov:rg}).
However, the class of spacetimes for which the description of bulk
position in field theory terms is understood is still very limited.
One of the goals of our paper is to attempt to extend the
understanding of this relation for the simplest examples of spacetimes
with a non-trivial causal structure.
 
In the relation (\ref{eq:uvir}), $U \to 0$ is mapped to diverging
scale size in the field theory. From the spacetime point of view, $U =
0$ in Poincar\'e coordinates is an event horizon, and one can think of
the divergence in the scale size as reflecting the one-way nature of
the horizon: particles at the event horizon cannot move to larger $U$,
and an infinite scale excitation can't return to smaller scale. As we
will review in section \ref{section:uvir}, at least in pure AdS, the
relation (\ref{eq:uvir}) provides a connection between spacetime and
field theory causality throughout spacetime~\cite{Kabat:causality}.
 
We would like to know if this connection between the UV/IR relation 
and causality can be generalised. A simple question to ask is whether 
the horizon of a black hole is also associated with an infinite scale 
size in the CFT. We will consider a variety of probes of black hole 
spacetimes, and find that the characteristic scale in the field theory 
description of time-independent probes is typically finite, even when 
they are very close to the horizon. Considering time-dependent probes 
is more complicated, but we argue that the scale size diverges at late 
times, although the leading behaviour is not directly related to the 
black hole structure. 
 
The example that we study is a charged black hole in AdS. Considering 
charged black holes allows us to have a large separation between the 
horizon size and the thermal scale. In an uncharged black hole, the 
standard relation (\ref{eq:uvir}), and some probe calculations, would 
assign a scale size which is of the order of the thermal scale when a 
probe is at the black hole horizon. We would like to investigate if 
this connection between the horizon and the thermal scale persists 
when there are other scales in the problem, or if we can see some sign 
of a divergent scale associated with the horizon. The presence of 
charge allows us to see which boundary scales are related to the 
thermal fluctuations in the gauge theory and which depend on the scale 
set by the black hole horizon. 
  
The black holes we study are the toroidal ``Reissner-Nordstr\"om AdS'' 
black holes.  In the case of AdS$_5$, these black holes can be 
obtained from the near horizon limit of spinning D3 
branes~\cite{Chamblin:charge1}.  Therefore the associated dual field 
theory is the world-volume theory on these branes. 
In~\cite{Chamblin:charge1, Chamblin:charge2}, the thermodynamic 
properties of charged Reissner-Nordstr\"om AdS$_{n+1}$ black holes was 
investigated. They considered both spherical black holes, with the 
boundary $\mathbf{R}\times S^{n-1}$, and toroidal black holes, where 
the asymptotic boundary is $\mathbf{R}^n$. From the point of view of 
thermodynamics, the spherical black holes are more interesting, but to 
analyse UV/IR relations, we will focus on the simpler case of toroidal 
black holes. These can be obtained as the infinite volume limit of the 
spherical black holes.  The metric for these black holes is 
\begin{equation} \label{eq:metric1}  
{ds}^2=-V(U){dt}^2+\frac{{dU}^2}{V(U)}+\frac{U^2}{R^2} 
\sum_{i=1}^{n-1}{dx}_i^2, 
\end{equation} 
where 
\begin{equation}     
V(U) = \frac{U^2}{R^2}-\frac{m}{U^{n-2}}+\frac{q^2}{U^{2n-4}}.  
\end{equation}  
We work in units where $l_s=1$, so $R = (g_{YM}^2 N)^{1/4}$.  
The horizon radius, $U_T$, is given by $V(U_T)=0$, and the temperature  
$T$ is related to the period $\beta$ in Euclidean time by 
\begin{equation}  
\beta=\frac{1}{T}=\frac{4\pi}{V'(U_T)}=\frac{4\pi  
R^2{U_T}^{2n-3}}{n{U_T}^{2n-2}-(n-2)q^2 R^2}.  
\end{equation} 
The black hole will be extremal ($T=0$, and the horizons coincide) at 
$U_T=U_e$, where ${U_e}^{2n-2}=(n-2)R^2 q^2/n$. It was shown 
in~\cite{Chamblin:charge1} that these black holes are 
thermodynamically stable (in both the canonical and the grand 
canonical ensembles) for arbitrary values of the mass and charge, so 
the black hole solutions carry information about the CFT in the 
corresponding ensemble. We will focus on the case of AdS$_5$, that is, 
$n=4$. 

It was further shown in~\cite{Hawking:charge} that the 
spherical black hole solutions in the AdS$_5 \times S^5$ context will 
not have superradiant modes, as the internal $S^5$ rotates at a speed 
less than the speed of light everywhere in the spacetime. It is easy 
to extend their argument to the toroidal black holes. The charged 
black hole (\ref{eq:metric1}) is derived from the reduction ansatz 
\begin{equation}  
ds^2=g_{\mu\nu}dx^{\mu}dx^{\nu}+\sum_{i=1}^3\left[{{d\mu}_i}^2+  
{\mu_{i}}^2\left(d{\phi}_i +A_{\mu}dx^{\mu}\right)^2\right]     
\end{equation}  
where $g_{\mu\nu}$ is the five-dimensional metric, ${\mu}_i$ are the  
direction cosines and ${\phi}_i$ are the rotation angles on the $S_5$.  
Non-zero $A_t$ gives the electric potential  
\begin{equation}  
A=(\Phi(U_T)-\Phi(U))dt, \textrm{  where  } \Phi(U)=\frac{q}{U^2}  
\end{equation}  
The norm of the Killing vector field $k=\partial/\partial t$ with  
respect to our ten dimensional metric is  
\begin{equation}  
k^2=-\left(1-\frac{{U_T}^2}{U^2}\right)\left[U^2   
\left(1-\frac{{U_-}^2}{U^2}\right) +  
\left(1-\frac{{U_-}^2}{{U_T}^2}\right)({U_T}^2+{U_-}^2)\right], 
\end{equation}  
where $U_-$ is the inner horizon.  We see that $k^2$ is always  
negative outside the black hole horizon, and thus superradiance cannot  
occur for the toroidal black hole.

A useful rewriting of the metric (\ref{eq:metric1}) is 
\begin{eqnarray} \label{eq:metric}  
ds^{2} = \frac{R^{2}}{U^{2}}\frac{dU^{2}}{f(U)} + 
\frac{U^{2}}{R^{2}}\left[-f(U)dt^{2}+\sum_{i=1}^3{dx_i}^{2}\right], \\ 
f(U) = 1 - (1+\theta)\frac{{U_T}^4}{U^4} + \theta\frac{{U_T}^6}{U^6}, 
\end{eqnarray}  
where we have defined a dimensionless parameter $\theta=q^2 R^2 / 
{U_T}^6$; for the uncharged black hole $\theta=0$, while the extremal 
black hole has $\theta=2$. The case $\theta=U_T=0$ is pure AdS$_5$ in 
Poincar\'e coordinates, where (\ref{eq:uvir}) is valid. 
  
After reviewing the connection between (\ref{eq:uvir}) and causality 
in section~\ref{section:uvir}, we move on to consider specific probes 
in this black hole background. We begin with a discussion of Wilson 
loops in section~\ref{section:loop}. We find that the qualitative 
behaviour of the loop expectation values is independent of the charge, 
and the non-trivial physics associated with the presence of a black 
hole appears at a scale given by (\ref{eq:uvir}). That is, the 
characteristic scale for these observables is the inverse horizon 
size, and is not in general related to the temperature. We then go on 
to discuss supergravity probes in section~\ref{section:sugra}. We find 
that the scale of the expectation value for time-dependent probes 
diverges at late times, but the expectation value is primarily 
determined by the asymptotic metric. We study the small contribution 
from the near-horizon region in the BTZ metric, and argue that it is 
spatially constant. For static supergravity probes, there is a finite 
scale size associated with the horizon in general.  The scale size 
associated with the static propagator diverges like $\ln(T)$ in the 
extremal limit $T \to 0$. This behaviour provides the main element of 
surprise in the paper, and it would be very interesting to gain a 
better understanding of this logarithmic dependence from the field 
theory point of view. 
  
In addition to addressing the implications for our understanding of 
bulk causality, we will briefly comment on the physical significance 
of these UV/IR relations from the field theory point of view, and 
remark on the relation to previous 
work~\cite{Csaki:glueball2,Russo:glueball} which studied the Euclidean 
rotating brane solutions as models for pure gauge theories. 
  
We conclude in section \ref{section:conclusions} with a discussion of 
the difficulties in identifying the origins of spacetime causal 
structure in the gauge theory, and some speculations for future 
directions. 
  
\section{UV/IR relations in black hole spacetimes} \label{section:uvir}  
  
To begin our investigation of the UV/IR relation in spacetimes with 
horizons, we review the connection between the UV/IR relation and 
causality in pure AdS proposed in~\cite{Kabat:causality}. In pure AdS, 
the condition for a bulk probe to move inside the light cone in the 
radial direction is 
\begin{equation} \label{eq:lc}  
- {U^2 \over R^2} dt^2 + {R^2 \over U^2} dU^2 \leq 0.  
\end{equation}  
For a supergravity probe, the UV/IR relation (\ref{eq:uvir}) implies 
that (\ref{eq:lc}) is equivalent to  
\begin{equation} \label{eq:kin}  
\left| {d \delta x_\parallel \over dt} \right| \leq 1 , 
\end{equation}  
which is just the statement that the field theory excitation's size 
cannot change faster than the speed of light in the field theory. This 
thus connects the causality of AdS to the causality of the space the 
field theory lives in.  
 
It would be interesting to see if a similar 
relation could exist for black hole spacetimes, through a suitable 
modification of the UV/IR relation (\ref{eq:uvir}).  
In a black hole spacetime, the condition (\ref{eq:lc}) 
is modified to 
\begin{equation} \label{eq:lc2}  
- {U^2 \over R^2} f(U) dt^2 + {R^2 \over U^2 f(U)} dU^2 \leq 0.  
\end{equation}  
We could derive a candidate UV/IR relation on the black hole spacetime 
by assuming that this spacetime condition is still equivalent to the 
kinematical condition (\ref{eq:kin}). If we assume a UV/IR 
relationship of the form $\delta x_{\parallel}= g(U)$, then 
(\ref{eq:kin}) would imply 
\begin{equation}  
\left\vert\frac{dU}{dt}\right\vert \le  
\left\vert\frac{dU}{dg(U)}\right\vert.   
\end{equation}  
Requiring that this condition is equivalent to (\ref{eq:lc2}) gives   
\begin{equation}  
\left\vert\frac{dg(U)}{dU}\right\vert=\frac{R^2}{U^2 f(U)}.   
\end{equation}  
This can be solved exactly for the black hole spacetimes we are 
considering, but the important feature is how the boundary scale size 
behaves for a probe near the black hole horizon.  The behaviour 
differs according to whether or not the black hole is extremal, but in 
both cases the scale size becomes infinite as the black hole horizon 
is approached. 
\begin{eqnarray} \label{eq:causal} 
\theta \ne 2 & \Rightarrow & \delta x_{\parallel} \sim  
\frac{R^2}{2(2-\theta)}\ln(U-U_T), \\ \theta = 2 & \Rightarrow & \delta  
x_{\parallel} \sim \frac{R^2}{12}\frac{1}{U-U_T}.   
\end{eqnarray}  
Thus, a diverging scale size for probes at finite radial position 
would be a signature of a horizon, under the assumption that bulk 
causality arises from a kinematical restriction in the field 
theory. This is the main prediction we will investigate in our 
discussion of the probes. 
  
In studies of the uncharged black holes, it was found that sources 
near the black hole horizon produced expectation values with a scale 
$\delta x_\parallel \sim 1/T$, where $T$ is the 
temperature~\cite{Banks:dynamics,Brandhuber:wilson1,Rey:wilson}. It 
was proposed that the physical interpretation of this is that probes 
which fall into the horizon become entangled with the thermal bath in 
the gauge theory. This suggests that the presence of 
finite-temperature Hawking radiation acts as a barrier to our ability 
to probe the non-trivial classical causal structure in the region near 
the horizon. One reason for our extension to the charged black holes 
is that it enables us to consider black holes of arbitrarily low 
temperature, and separate the thermal scale from the horizon radius, 
allowing us to study this issue further. 
 
\section{Wilson loop calculations} \label{section:loop} 
  
The expectation value of a Wilson loop operator in the field theory is
related by the AdS/CFT correspondence to the action for a string
worldsheet in the bulk which terminates on the path of the Wilson loop
on the boundary~\cite{Rey:wilson1,Maldacena:wilson}. This makes this a
very convenient observable to consider, as we can investigate its
leading-order behaviour by finding the minimal-area surface for the
string worldsheet. The symmetries of the metric (\ref{eq:metric}) lead
us to consider rectangular Wilson loops with two long sides, and a
separation $L$ between them. The long sides lie either along the $t$
direction (timelike Wilson loops) or along one of the $x^i$ directions
(spacelike Wilson loops).
  
\subsection{Timelike Wilson Loops}  
 
We wish to consider Wilson loop operators in the boundary theory,
where the loop $\mathcal{C}$ forms a rectangle with two long sides
extended along the $t$ direction, of length $t_0$, and two sides of
length $L$ along one of the $x^i$ directions. From the field theory
point of view, the value of this operator determines the potential of
an external quark-antiquark pair. This potential was obtained
in~\cite{Rey:wilson1,Maldacena:wilson} for the vacuum (pure AdS).  The
expectation value of the Wilson loop behaves like $\langle
W(\mathcal{C})\rangle\sim \exp(-t_0E)$ in the limit $t_0\to\infty$,
where $E=V(L)$ is the lowest possible energy of the quark-antiquark
configuration.  For large $N$ and large $g_{YM}^2 N$, this expectation
value is given by $\langle W(\mathcal{C})\rangle\sim\exp(-S)$, where
$S$ is the Nambu-Goto action for a fundamental string worldsheet which
joins the boundary at the loop $\mathcal{C}$ (we must also subtract
the infinite mass of the W-boson to regularize this supergravity
calculation). This calculation was extended to the field theory at
finite temperature by considering a string worldsheet in a
Schwarzschild-AdS background in~\cite{Rey:wilson,
Brandhuber:wilson1}\footnote{A review of Wilson loops from the
string/gauge correspondence can be found in
\cite{Sonnenschein:wilson1} which includes extensive references.}.  In
pure AdS, the quark-antiquark potential is $V \propto - (g^2_{YM}
N)^{1/2}/L$, as expected by conformal invariance. At finite
temperature, the small-$L$ behaviour is similar, but screening sets in
and $V =0$ at $L \sim 1/T$, where $T$ is the temperature.
 
\begin{figure}[p]  
\begin{center} 
\includegraphics[width=0.5\textwidth, height=0.25\textheight]{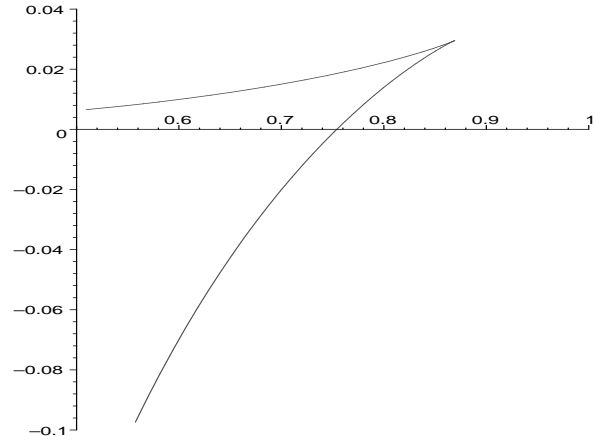} 
\end{center} 
\caption{$E$ vs.\ $L$ plot for uncharged black hole ($\theta=0$)}
\label{fig:1}  
\end{figure}

\begin{figure}[p]  
\begin{center} 
\includegraphics[width=0.5\textwidth, height=0.25\textheight]{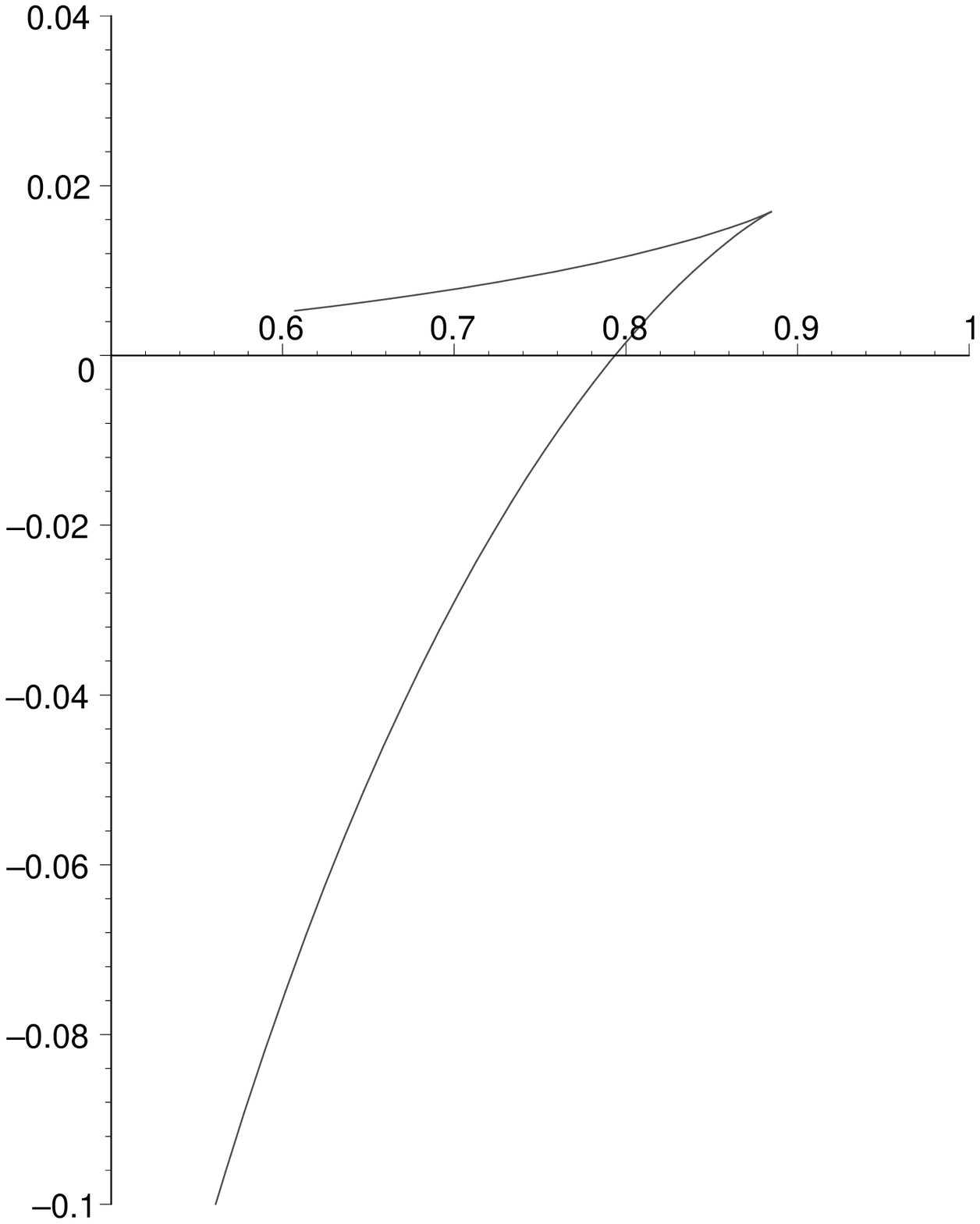} 
\end{center} 
\caption{$E$ vs.\ $L$ plot for charged black hole ($\theta=1$)}
\label{fig:2}   
\end{figure}

\begin{figure}[p]  
\begin{center} 
\includegraphics[width=0.5\textwidth, height=0.25\textheight]{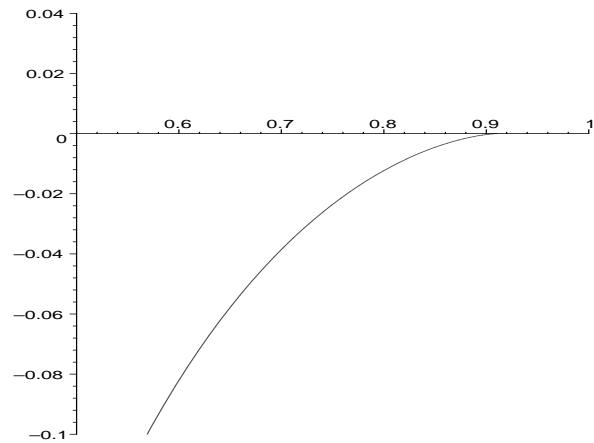} 
\end{center} 
\caption{$E$ vs.\ $L$ plot for extremal black hole ($\theta=2$)}
\label{fig:3}   
\end{figure}

Our objective is to investigate the quark-antiquark potential obtained 
from the string worldsheet in the charged AdS black hole background. 
We begin with the metric (\ref{eq:metric}), and calculate the string 
worldsheet area from the Nambu-Goto action 
\begin{equation} \label{eq:stract} 
S=\frac{t_0}{2\pi}\int dx\sqrt{(\partial_{x}U)^2+\frac{U^4}{R^4}f(U)}.  
\end{equation}  
We are working in the static gauge $\tau=t$, $\sigma=x$ for the 
string worldsheet coordinates, where $x$ is the position in the boundary 
direction along which the quarks are separated.  This action is 
independent of $x$, so we can calculate $x$ from the 
Euler-Lagrange equations to find 
\begin{equation}  
x=\frac{R^2}{U_T}\alpha\sqrt{1-(1+\theta){\alpha}^4+\theta{\alpha}^6}\int_1^{ 
  \frac{U\alpha}{U_T}}\frac{y^2 dy}{\sqrt{(y^2-1)(y^2-{\alpha}^2)(y^4+ 
  {\alpha}^2y^2-\theta{\alpha}^4)(y^4+y^2-\theta{\alpha}^6)}}.  
\end{equation}  
We have introduced a dimensionless parameter $\alpha=U_T / U_0$, where
$U_0$ is the minimum value of the radial coordinate along the string
worldsheet in AdS.  We see that $x=0$ at $U=U_0$, so the string
profile is symmetric about $x=0$.  Furthermore, the separation of the
quarks is given by $L=2x(U=\infty)$.  We wish to calculate the energy,
which is na\"{\i}vely given by $S/t_0$, where $S$ is the action
(\ref{eq:stract}) integrated over the range $-L/2\le x\le L/2$.
However, as was pointed out in \cite{Rey:wilson1,Maldacena:wilson} and
subsequent works, this would give an infinite result due to the
contribution from the mass of the W-boson.  We must therefore
regularize the expression by only integrating up to $U=U_{max}$, and
subtracting the regularized mass of the W-boson, $U_{max}/(2\pi)$.
Taking our cutoff to infinity, we then find the solution for the
energy
\begin{equation}  
E=\frac{U_T}{\pi\alpha}\left[\int_1^{\infty}\left(\frac{\sqrt{(y^2-{\alpha}^2) 
        (y^4+{\alpha}^2y^2-\theta{\alpha}^4)}}{\sqrt{(y^2-1)(y^4+y^2- 
        \theta{\alpha}^6)}} -1\right)dy-1+\alpha\right]. 
\end{equation}  
We can only evaluate this integral by numerical methods.  If we plot
$E/U_T$ against $L U_T/R^2$, the only free variable is $\theta$, which
specifies the charge on the black hole. We can study the effect of the
charge by varying $\theta$.  The results obtained in the uncharged
case have been reproduced here for comparative purposes in figure
\ref{fig:1}.  The typical behaviour of $E$ vs.\ $L$ for a charged
black hole (plotted here is the case $\theta=1$) is given in figure
\ref{fig:2}, and the behaviour for the extremal black hole is given in
figure \ref{fig:3}.
It should be noted that in these plots the parameter $\alpha$ is not 
plotted over the complete range, as numerical methods to solve the 
integral break down when $\alpha$ is close to $0$ or $1$.  However, 
the behaviour of the $E$ vs.\ $L$ plot in the unplotted regions can be 
observed from studying $L$ and $E$ separately as functions of 
$\alpha$.  As $\alpha\to 0$, $L\to 0$ and $E\to -\infty$.  With 
$\alpha$ increasing from $0$, the $E$ vs. $L$ plot rises smoothly to 
the cusp with increasing $L$ and increasing $E$, at which point both 
$L$ and $E$ begin to decrease along the upper branch.  This monotonic 
decreasing behaviour in $L$ and $E$ continues until $\alpha=1$ where 
both $L$ and $E$ are zero.  
 
In the uncharged case, it was remarked that the upper branch (with 
$E\ge 0$) is unphysical.  The potential energy shown in the graph is 
the energy of a U-shaped string configuration hanging into the bulk. 
There is an alternative configuration, a pair of strings hanging 
straight down to the horizon.  This second configuration has zero 
energy (after we subtract the W-boson mass contribution), so when the 
energy of the U-shaped configuration is greater than zero, this 
configuration is no longer energetically favourable and we pass over 
to the other.  We should therefore only consider the section of the 
$E$ vs.\ $L$ curves with negative energy.  Where the curve crosses the 
axis, the potential becomes constant. From the point of view of gauge 
theory, this corresponds to the screening of the quark charge by the 
plasma in the field theory which carries the energy of the thermal 
state. 
 
We see that the qualitative behaviour of the potential remains the 
same as the charge of the black hole is increased. The separation at 
which screening sets in increases slightly, but the overall scale is 
still set by the horizon radius $U_T$.  This is reasonable from the 
point of view of the field theory, since we interpret this screening 
as due to polarization in the plasma in the field theory which carries 
the energy density in this state (which corresponds to the black hole 
mass from the spacetime point of view). In the charged black hole 
case, this energy density goes like $U_T$, and not like the 
temperature. Even in the extremal, $T \to 0$ limit, there is still a 
finite energy density, which is responsible for the screening behaviour 
in figure~\ref{fig:3}. 
 
It is interesting to observe that the maximum value of the parameter 
$\alpha$ for which $E$ is negative increases from $\sim 0.66$ in the 
uncharged case to $1$ in the extremal charged case.  That is, as we 
increase the charge, the string worldsheet probes deeper into the 
interesting region near the horizon before the cross-over to the 
disconnected solution. Despite this behaviour, these loops are not a 
good probe of the bulk causality. In particular, there is no sign of 
any special behaviour as $T \to 0$. From the field theory point of 
view, the qualitative screening behaviour is associated with the 
background energy density, and one seems to find qualitatively similar 
behaviour independent of the details of the energy distribution. It is 
also possible to construct examples which display the same screening 
behaviour without a black hole horizon, for example by considering 
states on the Coulomb branch of the field 
theory~\cite{Freedman:coulomb,Brandhuber:coulomb,Giddings:shell}. 
Thus, while the value of the screening length is dictated by the 
horizon radius $U_T$, this probe is insensitive to the horizon as a 
horizon.

\subsection{Spacelike Wilson Loops}  
  
We can use similar techniques to study spacelike Wilson loops. 
Spacelike Wilson loops for the finite temperature field theory were 
considered in~\cite{Witten:thermal,Brandhuber:wilson2}.  In the 
uncharged case, one could consider the Euclidean AdS Schwarzschild 
black hole, and analytically continue one of the spatial $x^i$ 
directions instead of the time $t$. Because of the thermal boundary 
conditions, the $t$ direction is compactified on a circle of period 
$\beta=1/T$ in the Euclidean solution. At energies smaller than the 
compactification scale, we should obtain a pure gauge theory living in 
the $2+1$ uncompactified directions, as all the other modes of the 
original theory get a mass proportional to the temperature. The 
spacelike Wilson loop with the long side along the analytically 
continued direction is interpreted as giving the quark-antiquark 
potential in this gauge theory. The supergravity calculation indicated 
that it would display an area-law behaviour for $L\gg \beta$, in 
agreement with the expectation that this pure Yang-Mills theory is 
confining. Unlike the timelike loops, the string worldsheet always 
approaches arbitrarily close to the horizon as we increase the 
separation in the boundary.

In the charged case, the spacelike Wilson loop determined by string 
worldsheets in the metric (\ref{eq:metric}) does not have the same 
interpretation. To get a real Euclidean solution from 
(\ref{eq:metric}), we need to analytically continue both $t \to i\tau$ 
and $q \to i q'$. (This is easy to see if we remember that this charge 
comes from rotation in the higher-dimensional solution; that is, it is 
an angular momentum parameter.) This analytic continuation drastically 
changes the physics--for example, the analytically continued metric no 
longer has an extremal limit--and calculations of the spacelike Wilson 
loops in the Lorentzian metric (\ref{eq:metric}) are therefore not 
simply related to the properties of a $2+1$ pure gauge theory. Our 
motivation for studying the spacelike Wilson loops is thus simply that they 
are an interesting probe of the state in the $3+1$ supersymmetric 
field theory corresponding to our charged black holes.  
 
\begin{figure}[h]  
\begin{center} 
\includegraphics[width=0.5\textwidth,
height=0.2\textheight]{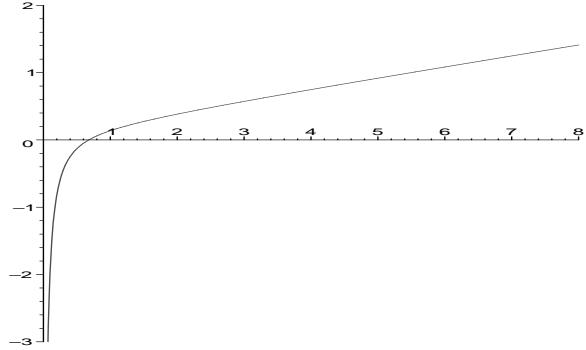}  
\end{center} 
\caption{Spacelike Wilson loop - $E$ vs.\ $L$ plot for extremal black 
hole. As before, we plot $E/U_T$ against $L U_T/R^2$.} \label{fig:4}  
\end{figure}

The action for the string worldsheet spanning a loop along two 
spatial dimensions is 
\begin{equation}  
S=\frac{Y}{2\pi}\int dx\sqrt{\frac{(\partial_{x}U)^2}{f(U)}+\frac{U^4}{R^4}}, 
\end{equation}  
where $Y$ is the length of the long side of the loop.  Repeating the 
method of calculation of $L$ and $E$ used above, we find 
\begin{eqnarray}  
L & = & \frac{2R^2\alpha}{U_T}\int_1^{\infty}\frac{y  
dy}{\sqrt{(y^4-1)(y^2-{\alpha}^2)(y^4+{\alpha}^2y^2-{\alpha}^4\theta)}},  
\\ E & = &  
\frac{U_T}{\pi\alpha}\left[\int_1^{\infty}\left(\frac{y^5}{
\sqrt{(y^4-1)(y^2-{\alpha}^2)(y^4+{\alpha}^2y^2-{\alpha}^4\theta)
}}-1\right)dy      
-1+\alpha\right].  
\end{eqnarray}  
We would like to find the behaviour of the theory at large $L$.  $L$ 
is increasing as a function of $\alpha$, so this requires us to 
consider the behaviour for $\alpha\to 1$ (i.e., we consider strings 
which hang close to the horizon).  For any value of the parameter 
$\theta$, both of the integrals are then dominated by the region 
$y=1$.  So for $\alpha\to 1$, the integrals in $L$ and $E$ become the 
same and we uncover the same area law as in the uncharged case: 
\begin{equation}  
E=\mathcal{T}L, 
\end{equation}  
where the tension $\mathcal{T}=\frac{U_{T}^2}{2\pi R^2}$. The $E$ vs.\ 
$L$ plot for the extremal black hole is given in figure (\ref{fig:4}), 
to illustrate the similarity with the uncharged case. The scale at 
which the linear behaviour sets in is once again determined primarily 
by the horizon radius $U_T$; the effect of $\theta$ is just some 
multiplicative factor of order unity. 
 
These probes see the horizon basically as a boundary, providing a 
lower bound on $g_{x^i x^i}$, and hence enforcing an area law 
behaviour at large distances. Since they don't probe the $g_{tt}$ part 
of the metric, it is not suprising that they're not good probes of the 
causal structure, or particularly sensitive to the temperature. 
  
\section{Supergravity probes} \label{section:sugra} 
 
We now consider the supergravity propagators on the charged black hole 
background. These can be used to calculate the dual expectation value 
for sources coupled to the supergravity fields near the horizon. One 
might hope that these supergravity sources will better probe the 
causal structure, as unlike the string worldsheets considered above, 
these sources can have compact support in the radial 
direction. However, the fact that the one-point functions are 
determined by the asymptotic fields will still complicate the story.   
  
\subsection{Retarded propagator}  
 
We consider first the retarded propagator, defined as the solution to 
the wave equation 
\begin{equation} 
\partial_\mu (\sqrt{-g} g^{\mu\nu} \partial_\nu 
G(x,x')) = \delta(x-x') 
\end{equation} 
subject to the boundary condition $G(x,x')= 0$ for $t<t'$. Here, $x'$ 
is the position of the source, and $x$ is the position where the 
measurement is made. It is natural to assume that sources in the bulk 
will follow geodesics, and in the charged black hole spacetimes, this 
implies that they will fall into the black hole. To study the effects 
of such source probes, we must calculate boundary expectation values 
using the retarded propagator; to learn about the causal structure, we 
must respect it. Unfortunately, explicit calculations are extremely 
difficult, except in $2+1$ dimensions, where the black hole is locally 
AdS. We will remark on the qualitative features of geodesic probes, 
assuming the solution has similar properties to the solution in pure 
AdS. 
 
The retarded propagator in pure AdS was previously investigated 
in~\cite{Danielsson:prop}, where an explicit solution was 
constructed. In AdS$_{2n+1}$, the retarded propagator is non-zero only 
in the part of the forward light cone that can be reached by a causal 
geodesic. In AdS$_{2n}$, it is non-zero only on the forward light 
cone. In either case, the retarded propagator from a point in the bulk 
to a point on the boundary is non-zero only where the forward light 
cone of the point in the bulk intersects the boundary. 
 
The expectation values dual to probes following geodesics in pure AdS 
were constructed in~\cite{Danielsson:prop}. There is an isometry which 
maps any geodesic in pure AdS to any other, which acts on the boundary 
as a conformal transformation. This can be used to obtain the 
expectation value dual to a boosted probe. These geodesics start from 
large radius in the Poincar\'e coordinates and fall towards the 
horizon. As the boost is increased, the initial position is moved to 
larger radius, bringing the probe closer to the boundary. It was found 
that as the boost is increased, the expectation value became 
concentrated in a `bubble' around the light-cone of the point where 
the source makes its closest approach to the boundary 
(see~\cite{Danielsson:prop} for details). An exact metric describing a 
lightlike particle falling from a point on the boundary of AdS was 
found in~\cite{Horowitz:particle}. In that case, the dual expectation 
value was a delta-function along the light cone of the point on the 
boundary that the particle fell from, in agreement with the general 
arguments for test particles. Thus, the expectation value spreads to 
infinite scale size as the particle falls towards the horizon. (Note 
that for a time-dependent probe, we have to wait for an infinitely 
long time to see this infinite scale size, which makes this a little 
inconvenient as a probe of causality.) 
 
Thus, most of the expectation value is along the light cone of the 
initial point when the source starts near the boundary. By causality, 
only the part of the source trajectory near the boundary can be 
contributing to this part of the expectation value; the region where 
the expectation value is large is outside of the light cone of all but 
the initial part of the probe's trajectory. Thus, in pure AdS, the 
expectation value lies mostly in an expanding bubble around the 
initial point, and is determined by the part of the source trajectory 
at large radial distances. 
 
This has several lessons for the black hole spacetimes. In black hole 
spacetimes, the metric at large distances is approximately AdS, so we 
would expect propagation in this region to be well-approximated by the 
propagation in pure AdS. Thus, for a source starting at large radial 
distance in the black hole geometry, the expectation value will have a 
contribution which produces a delta-function along the light 
cone. This part will spread out to infinite scale size along the light 
cone, just as it did in AdS. In an uncharged black hole background, 
there are thermal fluctuations around this average value, so it was 
argued in~\cite{Banks:dynamics} that in practice, we will see the the 
bubble expand until it reaches the thermal scale, where it becomes 
confused with the thermal fluctuations. In the charged black hole, we 
can suppress these thermal fluctuations, so we should be able to see 
the bubble expanding to larger and larger scales, just as in pure 
AdS.  
 
Although it is tempting to see this as a sign that the horizon is 
associated with infinite scale size, we should note that this 
expansion of the bubble is not connected with the horizon in the 
interior; it is determined by the behaviour of the probe far from the 
horizon. To see the effects of the non-trivial causal structure, we 
should consider the contribution to the expectation value from the 
region near the horizon. Because of the non-trivial causal structure, 
there can be a contribution from the region near the horizon only at 
very late times. Furthermore, the discussion in pure AdS suggests this 
contribution will be very small.  
 
If we consider the BTZ black hole in AdS$_3$, we can study the 
propagator from the near-horizon region, since the spacetime is 
locally pure AdS. The metric is~\cite{Banados:bh} 
\begin{equation} 
ds^2 = - {(r^2 - r_+^2) \over \ell^2} dt^2 + {\ell^2 dr^2 \over (r^2 - 
r_+^2)} + r^2 d\phi^2, 
\end{equation} 
where $\ell$ (the analogue of $R$ in our higher-dimensional 
discussion) is the cosmological length scale. If $\phi$ ranges over 
all values, this is a peculiar coordinate system for AdS$_3$. If 
$\phi$ is periodically identified with period $2\pi$, this is a black 
hole with a horizon at $r = r_+$. A lightlike geodesic starting at the 
boundary point $t = 0, \phi=0$ is described by 
\begin{equation} 
\phi = 0, r = r_+ \coth {tr_+ \over \ell^2}. 
\end{equation} 
The propagator from a point on this trajectory to the boundary at $r = 
\infty$ will be non-zero only at the intersection of the light-cone of 
the point with the boundary (since the space is locally AdS, we can 
use the propagator obtained in~\cite{Danielsson:prop}). For a point at 
$r= r_+ (1+\epsilon)$, the light cone meets the boundary at 
\begin{equation} 
t \approx {\ell^2 \over r_+} \ln \left( {2 \cosh(r_+ \phi/\ell) \over 
\epsilon} \right). 
\end{equation} 
As expected, $t$ here goes to infinity as the source approaches the 
horizon. More importantly, the point contributes at later times as we 
increase $\phi$. We consider the contribution to the expectation value 
at some fixed late time from the source worldline near the horizon. If 
we take the source sufficiently close to the horizon, we need only 
consider the contribution from the source at $\phi=0$, and not that of 
the images under the identification at $\phi = \pm 2\pi n$. That is, 
we can disregard the compactification of $\phi$ for this calculation. 
 
The expectation value for a geodesic source in these coordinates with 
$\phi \in (-\infty,\infty)$ is~\cite{Danielsson:prop} 
\begin{equation} \label{eq:exval} 
\langle {\cal O} \rangle = { (ar_+)^\Delta \over (a^2 + (1+a^2) 
\sinh^2 [r_+ (t+\phi)/2])^{\Delta/2} (a^2 + (1+a^2) 
\sinh^2 [r_+ (t-\phi)/2])^{\Delta/2}} 
\end{equation} 
for an operator of conformal dimension $\Delta$, where $a$ is the 
boost parameter (the source is lightlike for $a \to 0$). This 
expectation value is approximately independent of $\phi$ for $\phi \in 
(-\pi,\pi)$ at large $t$. Thus, the contribution to the expectation 
value from the region near the black hole horizon is 
$\phi$-independent, which is as close as we can come to infinite scale 
in the present context of a compact spatial direction on the boundary. 
We reiterate that this is just the contribution from the region near 
the horizon; the main contribution to the expectation value is at $t = 
\pm \phi$ as discussed in~\cite{Bal:particle}, and comes from the part 
of the worldline near the boundary. 
 
\subsection{Static propagator}  
  
We will next turn to static sources. The advantage of considering 
static sources is that since the source is always near the horizon, 
the expectation value will be affected by the near-horizon 
structure. However, the static propagator is independent of $g_{tt}$, 
so this is not guaranteed to produce a result which reflects the 
causal structure near the black hole, and in fact the answer we obtain 
does not have a straightforward relation to statements about the 
causality. However, it does appear to encode information about the 
near-horizon region in a non-trivial way. 
 
The calculation of the appropriate propagator in the uncharged black 
hole background was discussed in~\cite{Danielsson:prop}.  The static 
propagator for a massless scalar field is defined as a solution to the 
equation 
\begin{equation}  
\partial_{i}\left(\sqrt{g}g^{i j}\partial_{j}G(x,x')\right) =  
\sqrt{g_{tt}}\delta(x-x').  
\end{equation}  
Here $x'$ is the position of the source, while $x$ is the point at  
which the field is measured.  We take the metric   
(\ref{eq:metric}) and rescale the coordinates by $U\to U_T  
u$, $t\to t R^2 / U_T$ and $x\to x R^2 / U_T$. The equation for the  
static propagator is then  
\begin{eqnarray}  
\left(u^7-(1+\theta)u^3+\theta u\right)\partial_{u}^2\tilde{G} +  
\left(5 u^6-(1+\theta)u^2-\theta\right)\partial_{u}\tilde{G} - u^3 k^2  
\tilde{G} \nonumber\\ =  
\frac{\sqrt{u^6-(1+\theta)u^2+\theta}}{R^7}\delta(u-u').  
\end{eqnarray}  
Here we have Fourier transformed the propagator equation with respect  
to $x_i$, so   
\begin{equation}  
\tilde{G}(u,u',k_i) = \int d^3k e^{i\vec{k}\cdot\vec{x}}G(u,u',x_i).  
\end{equation}  
For $u<u'$ and $u>u'$ the Green's function is given by the solutions of  
the homogeneous equation  
\begin{equation} \label{eq:hom}  
\left(u^7-(1+\theta)u^3+\theta u\right)y''(u) + \left(5  
u^6-(1+\theta)u^2-\theta\right)y'(u) - u^3 k^2 y(u) = 0 . 
\end{equation}  
  
We must now consider the indicial equations which arise for solutions  
near the horizon $u=1$ and the boundary $u=\infty$.  For $u>u'$ this  
is $\sigma^2+4\sigma=0$, and the presence of electric charge makes no  
difference to the boundary behaviour of the Green's function.  In  
order for the solution to vanish at infinity, as required, we  
therefore have  
\begin{equation}  
\tilde{G}(u>u',k)=Ay_1(u,k) \textrm{ where }  
y_1(u,k)\sim\frac{1}{u^4}, u\to\infty.  
\end{equation}  
For the behaviour near $u=1$, we have the indicial  
equation $(4-2\theta)(\sigma(\sigma-1)+\sigma)=0$.  So long as   
$\theta\neq2$ (i.e., for a nonextremal black hole) this is also the same  
as in the uncharged case, and regularity at the horizon requires  
\begin{equation}  
\tilde{G}(u<u',k)=By_2(u,k) \textrm{ where } y_2(u,k)\sim 1, u\to 1.  
\end{equation}  
We will return to the extremal case later in this section.  The  
constants $A$ and $B$ are calculated by continuity in the Green's  
function at $u=u'$ and the correct discontinuity in its derivative, giving  
\begin{equation}  
A=\frac{y_2}{W(y_1,y_2)}\frac{1}{R^7
u'\sqrt{{u'}^6-(1+\theta){u'}^2+\theta}} ,
B=\frac{y_1}{W(y_1,y_2)}\frac{1}{R^7
u'\sqrt{{u'}^6-(1+\theta){u'}^2+\theta}}.
\end{equation}  
From (\ref{eq:hom}) we see that the the Wronskian $W(y_1,y_2)$ is 
\begin{equation}  
W(y_1,y_2)=\frac{w(k)}{u^5-(1+\theta)u+\frac{\theta}{u}}.  
\end{equation}  
The value of the dilaton can then be computed as 
in~\cite{Danielsson:prop}. The position-space propagator is written as 
the Fourier transform of this momentum-space propagator, and the 
integrals in this Fourier transform are expressed as a sum over the 
poles of $\tilde{G}(u,u',k_i)$. The boundary behaviour of the dilaton 
field is then found to be 
\begin{equation}  
\phi(U,x)\stackrel{U\to\infty}{\approx}\frac{{U_T}^4}{8\pi R^6  
U^4}\sum_{n=1}^{\infty}\frac{e^{-m_n r}}{m_n r w'(m_n)}\int  
du'y_2(u',m_n)du'.  
\end{equation}  
Here, $m_n$ are the zeros of $w(k)$, which give poles in the 
propagator. These correspond to the special values of $k$ for which we 
can construct solutions regular at both the horizon and infinity. 
$y_2(u',m_n)$ are the corresponding solutions of equation 
(\ref{eq:hom}).  Thus, the propagator has an exponential suppression 
for $r>1/m_n$, and these poles hence provide a maximum length scale 
for the expectation value dual to sources anywhere in the bulk. We 
proceed to determine this maximum scale by finding the poles $m_n$. 
 
In the case of the uncharged black hole, this problem of determining 
the zeros of the Wronskian physically corresponded to finding the 
glueball mass spectrum for the $2+1$ dimensional pure Yang-Mills 
theory, and was first described in~\cite{Witten:thermal}, and both 
numerical methods and analytic approximations have since been used to 
calculate the spectrum~\cite{Csaki:glueball, Minahan:glueball, 
Danielsson:prop}. As was emphasised in the discussion of spacelike 
Wilson loops, the Lorentzian metric (\ref{eq:metric}) is not directly 
related to the $2+1$ field theory obtained from a Euclidean rotating 
brane metric. Thus, the zeros $m_n$ found here will not be simply 
related to the glueball mass spectra obtained from studies of the 
rotating brane metrics in~\cite{Csaki:glueball2, Russo:glueball}. 
 
We follow the approach of \cite{Minahan:glueball} in calculating 
$m_n$, as the change of coordinates employed there makes the 
interpretation in the extremal limit clear.  Returning to equation 
(\ref{eq:hom}), with the change of variables $x=u^2$ ($k=i\kappa$), we 
find 
\begin{equation}  
\partial_{x}\left[(x^3-(1+\theta)x+\theta)\partial_{x}y\right]+{\kappa}^2
y=0.   
\end{equation}  
In order to use WKB methods on a second order linear differential 
equation, it is necessary to redefine the dependent variable so that 
it satisfies a differential equation with no first derivative term. 
The WKB analysis is greatly simplified with the change of variables 
$x=1+e^z$.  Defining  
\begin{equation}  
\psi=\sqrt{\frac{x^3-(1+\theta)x+\theta}{x-1}}y = \sqrt{f(z)}y,  
\end{equation} 
where  
\begin{equation} 
f(z)=e^{2z}+3e^z+(2-\theta),   
\end{equation}  
we obtain a differential equation which is completely analogous to the  
uncharged case,  
\begin{equation}  
\psi''+V(z)\psi=0,  
\end{equation} 
where  
\begin{equation} 
V(z)=\frac{{\kappa}^2}{f}e^z-\frac{f''}{2f}+\left(\frac{f'}{2f}\right)^2. 
\end{equation}  
The only change in this equation is that $f(z)$ is altered by the 
$\theta$ term.  To perform the WKB analysis we need to find the points 
where the potential in this equation is zero, as these are the turning 
points of the WKB approximation.  In the limits of large $\vert 
z\vert$, for $\theta \ne 2$, we have 
\begin{eqnarray} \label{eq:Vneg}  
V(z) & \approx &  
\left[\frac{\kappa^2}{2-\theta}-\frac{3}{2(2-\theta)}\right] e^z \textrm{  
for } z \ll 0, \\   
V(z) & \approx & {\kappa}^2 e^{-z}-1 \textrm{ for } z \gg 0.  
\end{eqnarray}  
For $\kappa$ sufficiently large there are thus turning points at  
$z=-\infty$ and $z=z_0 \approx 2\ln(\kappa)$.  The WKB approximation  
therefore gives  
\begin{equation}  
\left(n+\frac{1}{2}\right)\pi = \int_{-\infty}^{z_0}dz\sqrt{V(z)}.  
\end{equation}  
To leading order in $\kappa$ we can approximate the integral  
\begin{equation}  
\left(n+\frac{1}{2}\right)\pi=\int_{-\infty}^{\infty}\kappa  
\sqrt{\frac{e^z}{f(z)}}dz=\kappa  
\int_{1}^{\infty}\frac{dx}{\sqrt{x^3-(1+\theta)x+\theta}} \equiv 
\kappa \alpha, 
\end{equation}  
where the last equality defines $\alpha$. The zeros $m_n$ of the 
Wronskian are thus approximately given by 
\begin{equation}  
m_{n}=\frac{\pi}{\alpha}\left(n+\frac{1}{2}\right),  
\end{equation}  
where $n$ is a positive integer\footnote{It is argued 
in~\cite{Danielsson:prop} that the extra zero for $n=0$ does not 
contribute and this claim is substantiated by comparison to numerical 
results for calculation of the zeros in the uncharged black hole}. 
 
\begin{figure}[h]  
\begin{center} 
\includegraphics[width=0.5\textwidth, height=0.14\textheight]{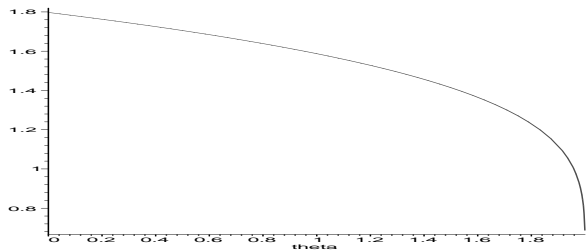} 
\end{center} 
\caption{$m_{1}$ vs.\ $\theta$} \label{fig:5} 
\end{figure}

In the uncharged case $\alpha$ can be evaluated exactly.  For the 
charged case, we evaluate alpha numerically, and see that as the 
charge of the black hole is increased, $\alpha$ increases and thus 
each $m_{n}$ decreases.  In figure~\ref{fig:5}, we plot the value of 
the lowest zero $m_{1}$ as a function of the parameter $\theta$ 
determining black hole charge.  As $\theta\to 2$, $\alpha$ diverges 
like $\ln(T) \propto \ln(2-\theta)$, where $T$ is the black hole 
temperature. Since we are obtaining a divergent answer, we should 
consider the validity of the approximation more carefully. 
  
The divergence found in the WKB approximation in the extremal case can 
be explained by considering the potential.  For non-extremal black 
holes the behaviour of $V(z)$ for $z \ll 0$ was as given in 
(\ref{eq:Vneg}).  This is the case for any value of $\theta$ other 
than $2$, but for $\theta=2$ it is the next term in $f(z)$ which 
contributes to $V(z)$ for $z \ll 0$.  The potential now no longer 
decays exponentially for $z \ll 0$: instead it becomes constant, and 
the potential never reaches zero in this region.  Therefore there is 
no second turning point of the equation and the bound state problem 
has no solutions. 
  
The different nature of the problem in the extremal case was 
discovered earlier, when we found for equation (\ref{eq:hom}), that 
the indicial equation for solutions near the black hole horizon is 
given by $(4-2\theta)(\sigma(\sigma-1)+\sigma)=0$.  This is true for 
$\theta\neq 2$, but for $\theta = 2$ the dominant terms in the 
solution near the horizon are those of a lower exponent and they lead 
to the indicial equation 
\begin{equation}  
12\sigma(\sigma-1)+24\sigma-k^2=0.  
\end{equation}  
For solutions to be well behaved near the horizon this requires 
$k^2\ge 0$.  This is problematic since the zeros are given by 
$m^2=-k^2$.  In fact, in the extremal case, (\ref{eq:hom}) can be 
solved exactly, enabling us to see how it differs from the 
non-extremal case. Making the substitution $x=u^2$ reduces the 
homogeneous equation for $\theta=2$ to 
\begin{equation}  
(x-1)^2 (x+2)\partial_{x}^2 y + 3(x-1)(x+1)\partial_{x}y -\frac{k^2}{4}y=0. 
\end{equation}  
Whereas in both the uncharged case and the nonextremal charged case  
our homogeneous differential equation was linear second order with  
four regular singularities, this equation only has three regular  
singularities, at $x=-2$, $x=1$ and $x=\infty$.  We recognise this as  
the hypergeometric equation and reduce it to the standard form by  
the transformation $z=3/(1-x)$. Then 
\begin{equation} \label{eq:hyper}  
z(1-z)\partial_{z}^2y -\partial_{z}y+\frac{k^2}{12}y=0.  
\end{equation}  
We will look for solutions of (\ref{eq:hyper}) satisfying our boundary 
conditions without restricting the sign of $k^2$, to see if any such 
solutions exist.  We need the solution to be normalizable, so 
$y\approx u^{-4}$ as $u\to \infty$; after coordinate transformations 
this condition becomes $y\approx z^2$ as $z\to 0$.  We also require 
the solution to be well-behaved at $u=1$, i.e., at $z=\infty$.  The 
hypergeometric equation has one solution with the correct behaviour at 
$z=0$, 
\begin{equation}  
y(z)=z^2 F\left(\frac{3}{2}+\lambda,\frac{3}{2}-\lambda,3;z\right),
\quad \lambda=\frac{1}{6}\sqrt{9+3k^2}\ge 0.   
\end{equation}  
To examine the behaviour near $z=\infty$, we use the asymptotic 
expansion in terms of hypergeometric functions in $1/z$  
to give 
\begin{eqnarray}  
y(z) & = &  
\frac{\Gamma(3)\Gamma(-2\lambda)}{{\Gamma(3/2-\lambda)}^2}(-1)^{
\frac{3}{2}+\lambda}z^{\frac{1}{2}-\lambda}F\left(\frac{3}{2}+
\lambda,-\frac{1}{2}+\lambda,1+2\lambda;\frac{1}{z}\right)   
\nonumber \\ & + &  
\frac{\Gamma(3)\Gamma(2\lambda)}{{\Gamma(3/2+\lambda)}^2}(-1)^{
\frac{3}{2}-\lambda}z^{\frac{1}{2}+\lambda}F\left(\frac{3}{2}-\lambda,
-\frac{1}{2}-\lambda,1-2\lambda;\frac{1}{z}\right).  
\end{eqnarray}  
Since the hypergeometric function as a function of $1/z$ is asymptotic 
to $1$ at $z=\infty$, for our function to be well-behaved at 
$z=\infty$ we require that $z$ does not appear outside the 
hypergeometric function with a positive exponent. In the second term 
of this expression for $y(z)$ this cannot be achieved, so the gamma 
function in the denominator must diverge to set this term to zero. 
However, this would only happen if $3/2+\lambda=-n$ for $n$ a 
non-negative integer, and $\lambda$ is positive.  Thus, in the 
extremal case there are no solutions of the time-independent wave 
equation satisfying the boundary conditions at both the horizon and 
the boundary. The breakdown in the WKB analysis near extremality is 
therefore physical. 
 
In these static propagator calculations, we have found that there is a 
finite screening length associated with most of the black hole 
spacetimes. From the calculations in the uncharged black hole, where 
the screening length is the thermal scale, one might have suspected 
that this is associated with the thermal fluctuations, which are 
concealing a divergence in the true behaviour. However, as we increase 
the charge, the screening length grows only logarithmically in the 
temperature, and soon falls below the thermal scale. There is thus 
really some limit on the characteristic scale for probes near the 
horizon, and we see no sign of a divergent scale size associated with 
the horizon in this calculation. It also would be interesting to 
understand the origin of this behaviour in the field theory: since the 
scale is not simply fixed by the horizon radius $U_T$, there may be 
some interesting physics here. From this point of view, it should be 
stressed that $1/m_n$ only provides an upper bound on the possible 
scale size; there may be power-law suppression at a smaller scale that 
this calculation is not sensitive to. 
  
\section{Conclusions} \label{section:conclusions}  
 
We only found signs of the expected divergence in the scale size of 
dual expectation values for probes near the horizon in the discussion 
of time-dependent probes. The failure to find such a relationship for 
Wilson loops is perhaps understandable, since the extended nature of 
the worldsheet implies that the part of the worldsheet that changes as 
we vary the asymptotic separation is probing a range of radii near the 
horizon, and not a specific value. Thus, the Wilson loops do depend 
non-trivially on the structure of the metric near the horizon, but 
don't really see the horizon as a horizon. 
 
It is more surprising that static supergravity probes produce dual 
expectation values with a finite scale size. From the spacetime point 
of view, this statement means that a point charge close to the horizon 
produces an asymptotic field which still depends non-trivially on the 
transverse coordinates $x_i$. This is quite different from the case of 
Schwarzschild black holes in flat space, where the field of a charged 
particle close to the horizon becomes completely spherically symmetric 
(see~\cite{macdonald:mem} and references therein). Note that this is 
not just the usual difference between the asymptotic behaviours of 
flat space and AdS: in the Schwarzschild case, the field measured at 
some finite radius is becoming spherically symmetric as the source 
approaches the horizon. It would be interesting to know what happens 
for the Schwarzschild-AdS solution. 
 
The static propagator also exhibits a mysterious logarithmic 
dependence on the temperature for small temperatures. It should be 
noted that this only provides an upper limit for the behaviour of the 
scale size of excitations for static probes near the horizon, and the 
actual scale could be constant. Nevertheless, it would be interesting 
to try to understand this behaviour from the field theory point of 
view. While no analogue of this behaviour was seen in the glueball 
mass calculations in rotating 
backgrounds~\cite{Csaki:glueball2,Russo:glueball}, this should not 
cause concern. As previously emphasised, these calculations address 
the physically different Euclidean solution obtained by $t \to i 
\tau$, $q \to i q'$. In this Euclidean solution, it is not possible to 
take the temperature to zero; in fact, the minimum value of the 
temperature is achieved when $q'=0$. 
  
One interesting example of a `static' source that is not covered by 
the foregoing analysis is a D-instanton. To consider a D-instanton, we 
need to pass to the Euclidean solution, so we cannot really think of 
it as a probe of the causal structure. However, the expectation value 
dual to a D-instanton is sensitive to the presence of the horizon in a 
dramatic way. For non-extremal black holes, the horizon in the 
Euclidean solution is simply a point where the proper length of the 
periodic $\tau$ direction goes to zero. If we place a D-instanton at 
this point, the translational symmetry in $\tau$ is preserved, so the 
dual expectation value must be $\tau$-independent. That is, for 
D-instantons near the horizon, the scale of the dual field theory 
instanton goes to infinity in the $\tau$ direction (and from the above 
comments on static propagators, in the $\tau$ direction only). This 
may be a useful test for a horizon in the Euclidean solution, but it 
does not help us to understand the causality of the Lorentzian 
solution. 
 
The key to a more satisfactory representation of the bulk causal 
structure may be to develop a relation between the bulk theory and the 
boundary which does not require us to propagate effects out to the 
boundary in spacetime. This is difficult to achieve with the present 
correspondence, as the relation between spacetime and field theory is 
phrased in terms of boundary conditions on the gravity side. It is 
worth stressing that this problem is distinct from the problem of 
studying local physics on scales smaller than the AdS scale; the black 
holes here can be as large as one wants.  Perhaps even resolving this 
apparently simple question requires the development of a more general 
background-independent version of the correspondence. 
 
\vskip.5in   
\centerline{\bf Acknowledgements}   
\medskip   
   
We are grateful for discussions with Ian Davies, Clifford Johnson, Esko
Keski-Vakkuri, Kenneth Lovis, Don Marolf, Antonio Padilla and David
Page. The work of JPG is supported in part by EPSRC studentship
9980045X. JPG thanks the Institut Henri Poincar\'e for hospitality
during the completion of this work.
  
\bibliographystyle{utphys}  
  
\bibliography{causal5}  
   
\end{document}